\magnification=\magstep1
\hfuzz=6pt
\baselineskip=15pt
\hsize=6.0truein
$ $
\vskip 1 in
\centerline{\bf Analog quantum error correction} 
\vskip 1cm
\centerline{Seth Lloyd$^1$}

\centerline{and}

\centerline{Jean-Jacques E. Slotine$^{1,2}$}
\bigskip
\centerline{1. d`Arbeloff Laboratory for Information 
Systems and Technology} 
\centerline{2. Nonlinear Systems Laboratory}
\centerline{Department of Mechanical Engineering}
\centerline{Massachusetts Institute of Technology}
\centerline{Cambridge, Massachusetts 02139}
\vskip 1cm
\noindent{\it Abstract:}  Quantum error-correction routines
are developed for continuous quantum variables such as
position and momentum.  The result of such analog quantum
error correction is the construction of composite continuous
quantum variables that are largely immune to the effects of noise and
decoherence.

\vskip 1.2cm
The quantum systems used for quantum computation and quantum
communications are small, sensitive, and easily perturbed [1-8]. 
The theory of quantum error-correcting codes provides a new
set of techniques for protecting quantum systems against the
effects of noise and decoherence [9-29].  Conventional quantum
error-correcting codes are only effective for discrete
variables, however.  This letter presents a set of {\it analog} quantum 
error-correcting routines that protect continuous variables
such as position and momentum against noise and decoherence.
These error-correcting routines can in principle be
enacted using simple Hamiltonian operations to stabilize
the states of arbitrary continuous quantum variables.  Particular
applications include error-correction for
quantum communications using continuous variables such as
photon momentum, and for analog quantum computers used
for simulating continuous quantum systems [30-31].

First consider the problem of correcting errors in a classical
discrete system.  The simplest binary error-correcting routine  
is triple modular redundancy, in which three bits are
initially set to the same value and checked at regular
intervals to see if they still have the same
value: if one of them differs, it is reset to the value
of the two others.  If the error rate per bit per unit
time is $\lambda$, then performing this `voting' routine
at intervals of time $ \delta t << 1/\lambda$ results in 
a new error rate of $3\lambda^2\delta t << \lambda$.

The discrete error-correcting technique of triple modular 
redundancy can be adapted simply to continuous classical variables.  
Consider three continuous variables $x_1x_2x_3$, initially
set to the same value $x$.  If at some brief time later
one of the three is found to differ from the other two,
it is reset to the majority value, $M$.  (If all three
differ, then $M$ can be taken to be the average value
of the two variables that differ the least.)  
The resetting can be accomplished by a simple nonlinear
dynamics such as $\ddot x_j = -k(x_j-M)-\gamma\dot x_j$,
where $\gamma, \sqrt{k}>>\lambda$. 
Just as in the discrete case,
if the probability of a variable being perturbed per unit
time is $\lambda$, then performing this nonlinear
`continuous voting' routine at intervals of time 
$ \delta t << 1/\lambda$ results in 
a new error rate of $3\lambda^2\delta t << \lambda$,
which can be made arbitrarily small by reducing
$\delta t$.

This classical continuous error-correcting routine
is clearly dissipative and decohering.  It can be
modified to preserve quantum coherence, however, and
can be used in a quantum context to protect
against some forms of quantum error.  Consider
three continuous `position' quantum variables with states
$|x_1x_2x_3\rangle_{123}$, and errors corresponding to
unitary operators $ e^{-iQ(P_j)}$, 
where $P_j = -i\partial/\partial x_j$ is the `momentum' operator
on the $j$'th variable and $Q$ is a polynomial function
of $P_j$ (we call these variables position and momentum for convenience
only: the method works for any continuous variable and
its conjugate).  Such an error takes
$$|x\rangle_i\rightarrow  
  e^{-iQ(P_j)} |x\rangle_j
=(1/\sqrt{2\pi})
\int_{-\infty}^\infty e^{-ipx-iQ(p)}|p\rangle_j {\rm d}p\eqno(1)$$
\noindent  where
$|p\rangle_j=(1/\sqrt{2\pi})
\int_{-\infty}^{\infty} e^{ipx}|x\rangle_j {\rm d}x$. 
The error acts on only one variable: 
$|x\rangle_k \rightarrow |x\rangle_k$ for
$k\neq j$.   For example, $Q(P_j) = \delta x P_j$
takes 
$$|x\rangle_j\rightarrow 
(1/\sqrt{2\pi})\int_{-\infty}^\infty e^{-ipx-ip\delta x}|p\rangle_j
{\rm d}p=|x+\delta x\rangle_j.\eqno(2)$$

To correct for these errors, apply the following
quantum `continuous voting' procedure. 
We assume that a variable
can be prepared in the state $|0\rangle_j$ by some
dissipative process such as cooling, and that the
state $|x\rangle_j$ can also be prepared, e.g., by applying
the `displacement' Hamiltonian $\eta x P_j$ to the state
$|0\rangle_j$ for a time $1/\eta$.  To `vote,' apply
the following procedure to three
continuous quantum variables, initially in the state
$|xxx\rangle_{123}$, together with three ancilla
variables $|x_1x_2x_3\rangle_{1'2'3'}$, initially 
in the state $|000\rangle_{1'2'3'}$:

\bigskip\noindent (0) 
Suppose that an error occurs to one
of the variables, e.g., the second one:
$$\eqalign{|x\rangle_2 &\rightarrow
 e^{-iQ(P_2)}|x\rangle_2\cr
&= (1/\sqrt{2\pi})
\int_{-\infty}^\infty e^{-ipx-iQ(p)}|p\rangle_2 {\rm d}p\cr
&\equiv \int_{-\infty}^{\infty} \alpha(x,x')|x'\rangle_2 
{\rm d}x'\cr} \eqno(3)$$
\noindent where $\alpha(x,x')= (1/\sqrt{2\pi})
\int_{-\infty}^\infty e^{-ip(x-x')-iQ(p)} {\rm d}p $.
Reprepare the ancilla variables
in the state $|000\rangle_{1'2'3'}$ (this corrects any
error that has occurred to the ancillae).  
The overall state of the variables and the ancillae is now
$$\big( |x\rangle_1 |0\rangle_{1'} \big)
\bigg(\int_{-\infty}^{\infty} \alpha(x,x')|x'\rangle_2{\rm d}x'
|0\rangle_2'\bigg) 
\big( |x\rangle_3|0\rangle_{3'}\big)\eqno(4)$$.

\noindent (1) Perform a continuous
quantum analog of voting.  We will assume that we can
perform simple real-number operations such as comparing the
values of two variables to see if they are equal, and
adding the value of one variable to another.  So
for example, we will assume that we can perform operations
such as comparing $|x_1\rangle_1$ and $|x_2\rangle_2$ to see if
$x_1=x_2$, and if they are, performing operations such as
$|x_1\rangle_1|x_2\rangle_2|x_3\rangle_3\rightarrow 
|x_1\rangle_1|x_2\rangle_2|x_3+x_1\rangle_3$. 
Such operations are reversible and correspond to
unitary transformations on Hilbert space.  They can
be accomplished by the application of simple interactions
between variables.  For example, the conditional addition 
operation just described can be accomplished by applying 
the Hamiltonian $\eta\Delta(X_1,X_2) X_1P_3$ for time 
$1/\eta$, where 
$$\Delta(X_1,X_2) 
= \int_{-\infty}^{\infty} |x\rangle_1\langle x| \otimes
|x\rangle_2\langle x| {\rm d}x.$$  
Such an operation
can be thought of as a continuous version
of a quantum logic gate.  (In real
life all such operations can be performed only to
finite precision; we will assume infinite precision
for the moment and discuss the effects of finite precision
below.)  If only one error
has occurred, then two of the $x$'s are always
equal.  So one by one, compare each of the $|x\rangle_i$
to the other two.   Let $ijk$ be some permutation of $123$.
If $x_i=x_j=x$,
then add the value of $x_k-x$ to the ancilla state 
$|0\rangle_k'$. If $x_i\neq x_j$ then do nothing.  
In our case, only $|x\rangle_2$ and its
ancilla will be affected.  The variables and ancillae
are now in the state 
 $$\big(|x\rangle_1|0\rangle_1'\big)
\bigg(\int_{-\infty}^{\infty} \alpha(x,x')|x'\rangle_2|x'-x\rangle_{2'}
{\rm d}x' \bigg)
\big(|x\rangle_3|0\rangle_{3'}).\eqno(5)$$

\noindent (2) Now if $x_i=x_j$ 
subtract the value of the $k$'th ancilla
variable from the original $k$'th variable, leaving the
state
$$\big(|x\rangle_1|0\rangle_{1'}\big)
\bigg( |x\rangle_2
\int_{-\infty}^{\infty} \alpha(x,x')|x'-x\rangle_{2'}
{\rm d}x'\bigg)
\big(|x\rangle_3|0\rangle_{3'}\big).\eqno(6)$$
\noindent Substituting in the explicit expression
for $\alpha(x,x')$ given above allows this state to be written
as $$\eqalign{&\big(|x\rangle_1|0\rangle_{1'}\big)
\bigg( |x\rangle_2
(1/2\pi)\int_{-\infty}^{\infty} e^{-ip(x-x')-iQ(p)}  |x'-x\rangle_{2'}
{\rm d}x' {\rm d}p \bigg)
\big(|x\rangle_3|0\rangle_{3'}\big)\cr
=& \big(|x\rangle_1|0\rangle_{1'}\big)
\bigg( |x\rangle_2
(1/\sqrt{2\pi})\int_{-\infty}^{\infty} e^{-iQ(p)}|p\rangle_{2'}
{\rm d}p \bigg)
\big(|x\rangle_3|0\rangle_{3'}\big).\cr
=& \big(|x\rangle_1 |0\rangle_{1'}\big)
\big( |x\rangle_2 e^{-iQ(P_{2'})}|0\rangle_{2'} \big)
\big(|x\rangle_3 |0\rangle_{3'}\big)\cr}
\eqno(7)$$
\noindent The error has now been corrected.

This procedure corrects the error by restoring
the three variables to the original continuous
`codeword' $|xxx\rangle_{123}$ while leaving the
ancilla in a state that is independent of the 
initial value of $x$.  The fact that the ancilla
is in a state that depends only on 
error operator $e^{-iQ(P_i)}$ applied and not on
the particular `codeword' to which it is applied means 
that the procedure restores not only continuous codewords
but arbitrary superpositions of the codewords 
$\int_{-\infty}^\infty \psi(x)|xxx\rangle{\rm d}x$. 

To continue correcting errors, simply return the ancilla
variables to $|000\rangle_{1'2'3'}$ 
and apply the procedure again a time $\delta t$ later. 
Just as in the 
classical case, performing the error-correcting routine 
at intervals $\delta t$ reduces the error rate 
from $\lambda$ to $3\lambda^2\delta t$, which can be
made as small as desired by decreasing $\delta t$.

It can easily be seen by interchanging the roles 
of $x$ and $p$ above that continuous codewords of the
form $|ppp\rangle_{123}$ can be protected against
arbitrary errors of the form $
e^{iR(X_j)}$ where $X_j$ is the position operator
on the $j$'th variable and $R$ is a polynomial
function of $X_j$.  In analogy to the $|xxx\rangle$ error-correcting
routine, we assume that variables and ancillae 
can be prepared in momentum eigenstates, $|p=0\rangle_{j},$
and that states $|p\rangle_j$ can be created by applying
the `boost' Hamiltonian $\eta p X_j$ to the state $|p=0\rangle_{j}$
for a time $1/\eta$.
The ancillae for the $|ppp\rangle$ error-correcting routine
a an be prepared initially in the state  $|p=0\rangle_{j'},$
or they can be prepared in position
eigenstates $|x=0\rangle_{j'}$ as before, and the interactions
between variables and ancillae adjusted to convert the value
of momentum in a variable to the value of position in the
ancilla. 

The following algorithm corrects both phase and displacement errors.
Define the state 
$$|{\bf p} \rangle_{123} \equiv (1/\sqrt{2\pi})
\int_{-\infty}^{\infty} e^{i{\bf p}x}|xxx\rangle_{123} {\rm d}x.\eqno(8)$$
\noindent  (Such a state can be created from the state
$$|{\bf p}\rangle_1|0\rangle_2|0\rangle_3 = 
(1/\sqrt{2\pi})
\int_{-\infty}^{\infty} e^{i{\bf p}x}|x\rangle_{1}
|0\rangle_2|0\rangle_3  {\rm d}x$$
\noindent by applying the Hamiltonian $\eta X_1P_j$ for
time $1/\eta$ to effect the unitary 
operation $|x\rangle_1 |y\rangle_j \rightarrow
|x\rangle_1|x+y\rangle_j$ for $j=2,3$.)
 
The error operator  $e^{iR(X_j)}$ has the same
effect on the triple-variable state $|{\bf p} \rangle_{123}$ 
that it has on the single-variable state $|p\rangle_j$:
$$|{\bf p}\rangle_{123}\rightarrow e^{iR(X_j)}|{\bf p}\rangle_{123}
=(1/\sqrt{2\pi}) 
\int_{-\infty}^{\infty} e^{i{\bf p}x + iR(x)}|xxx\rangle_{123}
{\rm d}x. \eqno(9)$$
\noindent This error can be corrected in an analogous way
to the errors on single continuous variables:
create redundant states of the nine variables 
$|{\bf p}_{123}{\bf p}_{456}{\bf p}_{789}\rangle_{1\dots 9}$ together
with a set of three ancilla variables originally in the state
$|000\rangle_{ABC}$, where ancilla variable $A$ is used
as the ancilla for the triple of variables $123$, $B$ is used
for $456$, and $C$ is used for $789$, then  
carry out the same error-correcting dynamics as above,
but as a function of the continuous variables ${\bf p}$ that
label the states $|{\bf p}\rangle$.
That is, phase errors on the triply-redundant state of 
triply-redundant continuous variables can be corrected 
by applying essentially the same error-correcting routine as before.

To correct any combination of phase and 
displacement errors on one variable, first apply
the $|xxx\rangle$ error-correction routine 
for error operators of the form $e^{-iQ(P_j)}$ to each
of the three triples of variables, $123,456,789$,
then apply the $|{\bf ppp}\rangle$ error-correction 
routine for error operators of the form $e^{iR(X_j)}$
to the nine variables as a whole.  
The basic idea of this continuous quantum error-correcting
routine is the same as Shor's binary quantum error
correcting routine [9]: using triple modular redundancy twice
(`triple-triple' modular redundancy) corrects both phase and 
displacement errors.
This sequence of error correcting steps compensates
for the effect of any error operator of the form
$e^{-iQ(X_j,P_j)}$, where $Q(X_j,P_j)$ is now a polynomial
in the operators $X_j,P_j$.  

To see the error-correction explicitly, use the
commutation relation $[X_j,P_j]=i$ to 
write $e^{-iQ(X_j,P_j)} = \sum_{m,n\geq 0} 
q_{mn}P_j^mX_j^n$.  Look at what happens
when an error of this form occurs to one
of the variables, for example, the first (j=1).
We have
$$\eqalign{|{\bf p}_{123}{\bf p}_{456}{\bf p}_{789}&\rangle_{1\dots 9} 
|0\ldots0\rangle_{1'\ldots 9'}|000\rangle_{ABC}\cr
\rightarrow 
\sum_{mn}q_{mn} &P_1^mX_1^n (1/\sqrt{2\pi})
\int_{-\infty}^{\infty} e^{i{\bf p}x}|xxx\rangle_{123}{\rm d} x\cr
&|{\bf p}\rangle_{456}|{\bf p}\rangle_{789}
|0\ldots0\rangle_{1'\ldots 9'}|000\rangle_{ABC}\cr}\eqno(10)$$
\noindent which can be rewritten using the decompositions
$|x\rangle = (1/\sqrt{2\pi})
\int_{-\infty}^{\infty} e^{-ipx}|p\rangle{\rm d}p$,
$|p\rangle = (1/\sqrt{2\pi})
\int_{-\infty}^{\infty} e^{i px'}|x'\rangle{\rm d}x'$,
as
$$\eqalign{ \sum_{mn} q_{mn}(1/\sqrt{2\pi})^3
&\int_{-\infty}^{\infty} p^mx^n 
 e^{i{\bf p}x}e^{-ip(x-x')}|x'\rangle_1|xx\rangle_{23}{\rm d}x
{\rm d}x' {\rm d}p\cr
&|{\bf p}\rangle_{456}|{\bf p}\rangle_{789}
|0\ldots0\rangle_{1'\ldots 9'}|000\rangle_{ABC}.\cr}\eqno(11)$$
\noindent Now proceed as before, comparing $x_i, x_j, x_k$,
and if $x_i=x_j=x$, adding $y=x'-x$ to the value of the 
ancilla state $|0\rangle_{k'}$ and
subtracting $x'-x$ from the value of the state $|x'\rangle_k$.
Only the first variable and its ancilla state will be
affected, resulting in the state
$$\eqalign{\sum_{mn} q_{mn}(1/\sqrt{2\pi})^3
&\int_{-\infty}^{\infty} p^mx^n
 e^{i{\bf p}x}e^{ipy}|y\rangle_{1'}|xxx\rangle_{123}{\rm d}x
{\rm d}y {\rm d}p\cr
&|{\bf p}\rangle_{456}|{\bf p}\rangle_{789}
|0\ldots0\rangle_{2'\ldots 9'}|000\rangle_{ABC}\cr
=\sum_{m,n}q_{mn}(1/\sqrt{2\pi})
&\int_{-\infty}^{\infty} x^n
e^{i{\bf p}x}|xxx\rangle_{123}{\rm d}x\cr
&|{\bf p}\rangle_{456}|{\bf p}\rangle_{789}
P^m_{1'}|0\rangle_{1'}|0\ldots0\rangle_{2'\ldots 9'}|000\rangle_{ABC},\cr
}\eqno(12)$$
\noindent where $P^m_{1'}$ acts only on the first
ancilla variable. 
The error-correction routine for states of
the form $|xxx\rangle$ has transferred the effect of
the $P_j^m$ part of the the error operator from
the codeword to the ancilla.

Similarly, applying the $|{\bf ppp}\rangle$
error-correction to the state in (12) transfers the
effect of the $X_j^n$ part of the error-operator
from the codeword to the ancilla, resulting in 
the state
$$\eqalign{(1/\sqrt{2\pi})
\int_{-\infty}^{\infty} 
 e^{i{\bf p}x}&|xxx\rangle_{123}{\rm d}x
|{\bf p}\rangle_{456}|{\bf p}\rangle_{789}\cr
&\big(\sum_{m,n} q_{mn}
P^m_{1'}|0\rangle_{1'} X^n_{A}|0\rangle_{A}\big)
|0\ldots0\rangle_{2'\ldots 9'}|00\rangle_{BC}\cr
=&|{\bf p}\rangle_{123}
|{\bf p}\rangle_{456}|{\bf p}\rangle_{789}
e^{-iQ(X_{A},P_{1'})}
|0\ldots0\rangle_{1'\ldots 9'}|000\rangle_{ABC}.\cr
}\eqno(13)$$
\noindent The error has now been corrected.  The ancillae
can be reset and the procedure repeated to provide
ongoing error correction.

To summarize: In each term of the polynomial
expansion of the error operator,
the application of $|xxx\rangle$ error-correcting
routine to the triple of continuous variables
containing $j$ restores the triple where the error
occurred to a superposition of the form 
$\int_{-\infty}^{\infty} \beta_n({\bf p, p'}) 
|{\bf p'}\rangle {\rm d}{\bf p'}$,
where $\beta_n({\bf p},{\bf p'})
=(1/\sqrt{2\pi}) \int_{-\infty}^{\infty}
x^ne^{i({\bf p}-{\bf p'})x}{\rm d}x$.
The subsequent application of the $|{\bf ppp}\rangle$ 
error-correction routine to the triple of
triples then restores the nine variables as a whole to the
state $|{\bf ppp}\rangle_{1\ldots 9}$.  The fact
that the state of the ancillae after each 
error-correcting routine depends only on
what errors occurred and not on which 
codeword $|{\bf ppp}\rangle_{1\ldots 9}$ the
system was in implies that
arbitrary superpositions of the form
$\int_{-\infty}^{\infty} \psi({\bf p})
|{\bf ppp}\rangle_{1\ldots 9}{\rm d}{\bf p}$ are also restored
by the continuous error-correction routine.  

The analog quantum error-correcting routine presented
above corrects for errors that are arbitrary polynomials
in $X_j$ and $P_j$.  It can be enacted in principle using simple
operations on the real numbers such as comparing and adding
two numbers.  What happens when these operations can
only be performed to finite precision?  By going
through the error-correcting
routine and following what happens when comparison and
addition are performed to finite precision $\delta$,
one can verify that the procedure still works as long
as (i) the wave-function $\psi$ does not vary significantly
over scales $\delta$, and (ii) the expectation
values for the error operators on the range of $\psi$
do not vary significantly over scales $\delta$.
Perhaps the easiest way to see why such inexact
error-correction still works 
is to note that when (i-ii) hold for finite precision $\delta$ in manipulations
of continuous variables the system behaves like an infinite-dimensional
{\it discrete} system with states $|x_n\rangle = |n\delta\rangle$.
The continuous error-correcting scheme above, performed
at finite precision, still functions as an error-correcting 
scheme for the discrete infinite-dimensional system. 
Similarly, the method described here generalizes in a straightforward
fashion to systems that are continuous in one variable
and discrete in the complementary variable (e.g., a particle
in a box).  

Our method is continuous but time-dependent:
it may be possible to devise error-correcting 
dynamics that are time-independent as well.
A particularly interesting application of analog quantum
error correction is to a collection of systems that evolve
according to a master equation:
$$\dot\rho= -i[H,\rho] + \sum_m \big( L_m\rho L_m^\dagger
-(1/2)L_m^\dagger L_m \rho -(1/2) \rho L_m^\dagger L_m
\big) \eqno(14)$$
\noindent Suppose that each $L_m$ is a polynomial function of
position and momentum operators acting on individual
subsystems (i.e., each subsystem sees an effectively
distinct environment; when the particles are coupled to
the same environment it may be possible to use symmetrized
states of the particles to resist noise and decoherence [32]).  
This is the typical case of particles
each governed by a distinct single particle master 
equation such as the optical master equation.
Inducing the proper interactions with ancillary 
continuous systems and applying the error-correcting routine
given above at intervals $\delta t$
allows nine subsystems to be grouped
together into one composite system whose states
$\int_{-\infty}^{\infty} \psi({\bf p})
|{\bf ppp}\rangle_{1\ldots 9}{\rm d}{\bf p}$ are unaffected by the 
dynamics (14) to first order in $\delta t$. 
For the continuous error-correction to be effective,
it must be repeated at intervals shorter than the dynamic
time-scales of the system such as its decoherence
time or spontaneous emission time.
The analog quantum error-correcting routine presented
here allows the creation of joint states of 
a composite continuous system that are largely
immune to the effects of interference and noise in principle.
In practice, of course, performing the continuous `quantum logic
gates' necessary to enact the analog error-correcting
scheme is likely to prove difficult. 

We have presented a quantum error-correcting
routine for continuous variables.  The routine allows
the creation of states of a composite system that
resist the effects of errors and noise.  For simplicity
of exposition, we presented a method for analog quantum
error correction based on Shor's original error-correcting
routine for qubits.  A variety of other continuous quantum
error-correcting routines can be constructed based on other 
discrete quantum codes.  In particular, in analogy to
[29], it should be
possible to devise a `perfect' analog quantum error-correcting
code using only five continuous variables, although the
dynamics of the error correction will be more complicated
than the simple continuous voting used here [33].   
The quantum error-correcting mechanism described here
is an example of a feedback loop that preserves
quantum coherence as proposed by Lloyd [34].  The
nonlinear dynamics cause the ancilla variables to
become correlated with the system in a coherent
manner, and the information that they possess is
used coherently to restore the system to its desired state.

\vfill
{\it Acknowledgements:} This work was supported by ONR and by
DARPA/ARO under the Quantum Information and Computation
initiative (QUIC).  
\eject
\centerline{\it References}
\bigskip

\noindent 1. R. Landauer, {\it Nature}, Vol.
{\bf 335}, pp. 779-784 (1988).
 
\noindent 2. S. Lloyd, {\it Science}, Vol. {\bf 263}, p. 695 (1994).
 
\noindent 3. W.G. Unruh, 
{\it Physical Review A}, Vol. {\bf 51}, pp. 992-997
(1995).
 
\noindent 4. I.L. Chuang, R. Laflamme, P.W. Shor, W.H. Zurek,
{\it Science}, Vol. {\bf 270}, pp. 1633-1635 (1995).

\noindent 5. R. Landauer, 
{\it Physics Letters A}, Vol. 217, pp. 188-193 (1996).
 
\noindent 6. R. Landauer, 
{\it Philosophical Transactions of the Royal Society
of London A}, Vol. {\bf 335}, pp. 367-376 (1995).
 
\noindent 7. G.M. Palma, K.-A. Suominen, A.K. Ekert,
{\it Proceedings
of the Royal Society A}, Vol. {\bf 452}, pp. 567-584 (1996).
 
\noindent 8. W.H. Zurek, {\it Physical Review Letters},
Vol  {\bf 53}, pp. 391-394 (1984).
 
\noindent 9. P.W. Shor, {\it Physical Review A}, Vol.
{\bf 52}, pp. R2493-R2496 (1995).
 
\noindent 10. A.M. Steane, {\it Physical Review Letters},
Vol. 77, pp. 793-797 (1996).
 
\noindent 11. A.R. Calderbank and P.W. Shor, 
{\it Physical Review A}, Vol. {\bf 54}, pp. 1098-1106 (1996).
 
\noindent 12. R. Laflamme, C. Miquel, J.P. Paz, W.H.
Zurek, {\it Physical Review Letters}, Vol. 77, pp. 198-201 (1996).
 
\noindent 13. P. Shor, 
{\it Proceedings of the 37th Annual Symposium on the Foundations
of Computer Science,} IEEE Computer Society Press, Los Alamitos,
1996, pp. 56-65.
 
\noindent 14. D.P. DiVincenzo and P.W. Shor, 
{\it Physical Review Letters}, Vol. {\bf 77}, pp. 3260-3263
(1996).
 
\noindent 15. J.I. Cirac, T. Pellizzari, P. Zoller, 
{\it Science}, Vol. {\bf 273}, 1207-1210 (1996).
 
\noindent 16. E. Knill and R. Laflamme, {\it Physical Review A}, Vol.
{\bf 55}, pp. 900-911 (1997).
 
\noindent 17. A. Steane, {\it Proceedings of the
Royal Society of London A}, Vol. {\bf 452}, pp. 2551-2577 (1996).
 
\noindent 18. C.H. Bennett, D.P. DiVincenzo, J.A. Smolin, W.K.
Wootters, 
{\it Physical Review A}, Vol. {\bf 54}, pp. 3824-3851 (1996).
 
\noindent 19. B. Schumacher and M.A. Nielsen, {\it Physical Review
A}, Vol. {\bf 54}, pp. 2629-2635 (1996).
 
\noindent 20. B. Schumacher, 
{\it Physical Review A}, Vol.  {\bf 54}, pp. 2614-2628 (1996).
 
\noindent 21. C.H. Bennett, G. Brassard, S. Popescu, B. Schumacher,
J.A. Smolin, W.K. Wootters, {\it
Physical Review Letters}, Vol. {bf 76}, pp. 722-725 (1996).
 
\noindent 22. A. Ekert and C. Macchiavello, 
{\it Physical Review Letters}, Vol. {\bf
77}, pp. 2585-2588 (1996).
 
\noindent 23. S. Lloyd, 
{\it Physical Review A}, Vol. {\bf 55}, pp. 1613-1622 (1997).
 
\noindent 24. J.I. Cirac, P. Zoller, H.J. Kimble, H. Mabuchi,
{\it Physical Review Letters}, Vol. {\bf 78}, pp. 3221-3224 (1997).
 
\noindent 25. C.H. Bennett, D.P. DiVincenzo, J.A. Smolin,
{\it Physical Review Letters}, Vol. {\bf 78}, pp. 3217-3220 (1997).
 
\noindent 26. S.J. van Enk, J.I. Cirac, P. Zoller, {\it Physical
Review Letters}, Vol. {\bf 78}, pp. 4293-4296 (1997).

\noindent 27. D. Gottesman, {\it Phys. Rev. A}
{\bf 54}, 1862 (1996).

\noindent 28. H.F. Chau, {\it Phys. Rev. A}
{\bf 55}, 839 (1997).

\noindent 29. H.F. Chau, {\it Phys. Rev. A}
{\bf 56}, R1 (1997).

\noindent 30. R.P. Feynman, 
{\it International Journal of Theoretical Physics},
Vol. {\bf 21}, pp. 467-488 (1982).

\noindent 31. S. Lloyd, {\it
Science}, Vol. {\bf 273}, pp. 1073-1078 (1996).

\noindent 32. P. Zanardi and M. Rasetti, `Noiseless
Quantum Codes,' submitted to {\it Phys. Rev. Lett.}.

\noindent 33. S. Lloyd and J.-J.E. Slotine, to be published.
 
\noindent 34. S. Lloyd, `Controllability and Observability
of Quantum Systems,' submitted to {\it Phys. Rev. A}.

\vfill\eject\end